\documentclass[a4paper,reqno,10pt]{amsart}
\usepackage[all]{xy} 
\usepackage{amssymb} 
\usepackage{hyperref}
\usepackage{eucal}

\usepackage[all]{xy} 

\def\qed{\ifvmode\removelastskip\fi
{\unskip\nobreak\hfil\penalty50\hbox{}\nobreak\hfil \hbox{\vrule
height1.2ex width1.2ex}\parfillskip=0pt \finalhyphendemerits=0
\par\smallskip}}

        \author[N. Rom\'an-Roy]{N. Rom\'an-Roy}
        \address{Narciso Rom\'an-Roy:
        Departamento de Matem\'{a}tica Aplicada IV, Edificio C-3, Campus Norte
         UPC,\\
        C/ Jordi Girona 1, E-08034 Barcelona, Spain} \email{nrr@ma4.upc.edu}
        \author[M. Salgado]{M. Salgado}
        \address{Modesto Salgado:
        Departamento de Xeometr\'{\i}a e Topolox\'{\i}a, Facultade de
        Matem\'{a}ticas,
         Universidade de Santiago de Compostela,
         15782-Santiago de Compostela, Spain}
        \email{modesto@zmat.usc.es}

        \author[S. Vilari\~no]{S. Vilari\~no}
        \address{Silvia Vilari\~no:
        Departamento de Xeometr\'{\i}a e Topolox\'{\i}a, Facultade de
        Matem\'{a}ticas,
        Universidade de Santiago de Compostela,
        15782-Santiago de Compostela, Spain}
        \email{silvia.vilarino@usc.es}
        \title[sopdes  and nonlinear connections]
      {sopdes  and nonlinear connections}

        \keywords{{\sc sopde}, Nonlinear connection.}

        \subjclass[2000]{53C05,70S05}

\begin{document}

        \begin{abstract}

        The canonical $k$-tangent structure on $T^1_kQ=TQ\oplus\stackrel{k}{\ldots}\oplus TQ$ allows us to characterize nonlinear connections on $T^1_kQ$ and to develop  G\"{u}nther's ($k$-symplectic) Lagrangian formalism. We study the relationship between  nonlinear connections and second-order partial differential equ\-ations ({\sc sopde}s), which appear in  G\"{u}nther's Lagrangian formalism.
        \end{abstract}

        \maketitle

\begin{center}
[Published in {\sl Publ. Math. Debrecen} {\bf 78/2} (2011), 297Ð316 DOI: 10.5486/PMD.2011.4631]
\end{center}
\bigskip\bigskip

\theoremstyle{plain}
\newtheorem{prop}{Proposition}
\newtheorem{thm}{Theorem}[section]
\newtheorem{lem}[thm]{Lemma}
\newtheorem{state}{State}[section]
\theoremstyle{definition}
\newtheorem{definition}{Definition}[section]
\theoremstyle{remark}
\newtheorem{remark}{Remark}[section]
\newtheorem{example}{Example}

\newenvironment{proof*}[1][\textsc{\proofname}]{\par
  \pushQED{}%
  \normalfont \topsep6\p@\@plus6\p@\relax
  \trivlist
  \item[\hskip\labelsep
        \indent
        \scshape
    #1\@addpunct{.}]\ignorespaces
}{%
  \popQED\endtrivlist\@endpefalse
}
\newenvironment{pf}{\proof[\proofname]}{\popQED\endtrivlist}

\def\derpar#1#2{\frac{\partial{#1}}{\partial{#2}}}
\def\d{{\rm d}}
\def\tkq{{T^1_kQ}}
\def\taukq{{\tau_Q^k}}
\def\r{{\mathbb{R}}}
\def\rk{{\mathbb{R}^k}}
\def\inn{\mathop{i}\nolimits}
\def\Cinfty{{\rm C}^\infty}

\section{Introduction}\label{section 1}

Lagrangian Mechanics have been entirely geometrized in terms of symplectic geometry. In this approach there exists certain dynamical vector field on the tangent bundle of a manifold whose integral curves are the solutions of the Euler-Lagrange equations. This vector field is usually called {\it second-order differential equation} ({\sc sode} to short) or {\it spray} (sometimes it is called {\it semispray} and the term spray is reserved to homogeneous second-order differential equations, see for instance, \cite{grif1,LR-1988}). Let us remember that a {\sc sode} on $TQ$ is a vector field on $TQ$ such that $JS=C$, where $J$ is the {\it almost tangent structure} or {\it vertical endomorphism} and $C$ is the {\it canonical field} or {\it Liouville field}.

In \cite{grif1,grif2,grif3}, Grifone studies the relationship among {\sc sode}s, nonlinear connections and the autonomous Lagrangian formalism. This study was extended to the non-autonomous case by
M. de Le\'{o}n and P. Rodrigues \cite{LR-1988}.

The natural generalization to Classical Field Theory of the concept of {\sc sode} is called {\it second order partial differential equation} ({\sc sopde} to short). This concept was introduced by G\"{u}nther in \cite{GU-1987} in order to develop his Lagrangian polysymplectic (k-symplectic) formalism. The
  ``phase space" of this formalism is the tangent bundle of $k^1$-velocites $T^1_kQ$, that is, the Whitney sum of $k$-copies of the tangent bundle $TQ$,
$$T^1_kQ:= TQ \oplus \stackrel{k}{\ldots} \oplus TQ\,.$$

In this paper we study the relationship between nonlinear connections and arbitrary {\sc sopde}s on $T^1_kQ$.

The structure of the paper is the following:

In section \ref{cses} we describe briefly  the   tangent bundle of
$k^1$-velocities $\tkq$ of a manifold $Q$ (see  \cite{Morimoto-1970}). After, following to Griffone \cite{grif1,grif2,{grif3}} and Szilasi \cite{Szilasi} we define a canonical short exact sequence
\[\xymatrix{0\ar[r] & \tkq\times_Q\tkq\ar[r]^-{\bf i}
&T(\tkq)\ar[r]^-{\bf j} & \tkq\times_Q TQ\ar[r] &0}\]
wich allows us to introduce in an alternative way the canonical geometric elements on $T^1_kQ$: {\it the Liouville vector field} and {\it the canonical $k$-tangent structure}. The usual definition of these geometric elements can be found in \cite{MRS-2004,RRS-2005,RSV-2007(2)}.

In section \ref{Nc} we give two characterizations of the nonlinear connections on $\taukq:\tkq\to Q$. In the first one we use the canonical short exact sequence constructed in section \ref{cses} in an analogous way to that one   in Szilasi's book \cite{Szilasi} for the case $k=1$. In the second one we characterize nonlinear connections on $\taukq:\tkq\to Q$ using the canonical $k$-tangent structure $(J^1,\ldots, J^k)$. In the particular case $k=1$ we reobtain some results given by Grifone in \cite{grif1,grif2,grif3}.

Finally in section \ref{sopde} we recall the notion of  {\sc sopde}s (second order partial differential equations) and we study the relationship between {\sc sopde}s and nonlinear connections on $T^1_kQ$.

Along the paper we have used the Szilasi's HandBook \cite{Szilasi} and Griffone's papers \cite{grif1,grif3}
as principal reference.

All manifolds are real, paracompact, connected and $C^\infty$. All
 maps are $C^\infty$. Sum over crossed repeated indices is understood.

\section{The canonical short exact sequence.}\protect\label{cses}

In this section we describe briefly  the   tangent bundle of
$k^1$-velocities $\tkq$ of a manifold $Q$ (see  \cite{Morimoto-1970}), that is, the Whitney sum of $k$-copies of the tangent bundle $TQ$, which is the phase space where the $k$-symplectic Lagrangian formalism of classical field theories (G\"{u}nther's formalism \cite{GU-1987}) is developed. After, following to Griffone \cite{grif1,grif2,{grif3}} and Szilasi \cite{Szilasi} we define a canonical short exact sequence which allows us to introduce  the canonical geometric elements on $T^1_kQ$, which are necessary to develop the $k$-symplectic Lagrangian formalism: {\it the Liouville vector field and the canonical $k$-tangent structure}.

Moreover, the canonical short exact sequence introduced in this section will be used, in the following section, to characterized nonlinear connections on $T^1_kQ$.

\bigskip

\noindent $\bullet${\bf The tangent bundle of $k^1$-velocities of a
manifold.}\label{section 2}

\bigskip

 Let $Q$ be a $n$-dimensional differential manifold,
and let $\tau_Q:TQ\to Q$ be the tangent bundle of $Q$.
Denote by $T^1_kQ$ the Whitney sum $TQ \oplus
\stackrel{k}{\dots} \oplus TQ$ of $k$ copies of $TQ$, with
projection $\tau^k_Q : T^1_kQ \to Q$, $\tau^k_Q
({v_1}_q,\ldots,{v_k}_q)=q$. The fibre on
$q\in Q$ is the $nk$-dimensional vector space
$(T^1_kQ)_{q}=T_{q}Q\oplus\stackrel{k}{\ldots}\oplus
T_{q}Q$ . Along this paper an element of $T^1_kQ$ will be denoted by $\mathbf{v}_q=({v_1}_q,\ldots,{v_k}_q)$.

 The manifold $J^1_\mathbf{0}(\r^k,Q)$ of  $1$-jets of maps with source at $\mathbf{0}\in \r^k$ and projection map $\tau^k_Q: J^1_\mathbf{0}(\r^k,Q) \to Q$, $\tau^k_Q
(j^1_{\mathbf{0},q}\sigma)=\sigma (\mathbf{0})=q$, can be identified with $\tkq$ as
follows
\[
\begin{array}{ccc}
J^1_\mathbf{0}(\r^k,Q) & \equiv & TQ \oplus \stackrel{k}{\dots} \oplus TQ \\
j^1_{\mathbf{0},q}\sigma & \equiv & ({v_1}_q,\ldots,{v_k}_q)
\end{array}
\]
where $q=\sigma (\mathbf{0})$,  and ${v_A}_q=
\sigma_*(\mathbf{0})(\frac{\partial}{\partial t^A}(\mathbf{0}))$.
 $T^1_kQ$  is called the {\it tangent bundle of  $k^1$-velocities} of $Q$, see
\cite{Morimoto-1970}.

If $(q^i)$ are local coordinates on $U \subseteq Q$, then the induced
local coordinates   $(q^i , v_A^i)$ on $T^1_kU=(\tau^k_Q)^{-1}(U)$ are given by
$$q^i(\mathbf{v}_q)= q^i({v_1}_q,\ldots,{v_k}_q)=q^i(q),\qquad
  v^i_A(\mathbf{v}_q)= v_A^i({v_1}_q,\ldots,{v_k}_q)={v_A}_q(q^i) \, .$$

\bigskip

\noindent $\bullet$ \noindent{\bf The vector bundle $(\tkq\times_Q
\tkq,(\tau^k_Q)^*\tau^k_Q,\tkq)$.}

\bigskip
Let us consider the fiber bundle $\tau^k_Q:\tkq\to Q$ and the pull-back bundle of  $\taukq$ by $\taukq$, that is,
$$(\tkq\times_Q \tkq,(\tau^k_Q)^*\tau^k_Q,\tkq)\,,$$ where the
 total space is the fibre product
$$\tkq\times_Q \tkq=
\{(\mathbf{v}_{q},\mathbf{w}_{q})\in \tkq\times \tkq\;|
\; \tau^k_Q(\mathbf{v}_{q})=\tau^k_Q(\mathbf{w}_{q})\}\;, $$
and
$(\tau^k_Q)^*\tau^k_Q:\tkq\times_Q\tkq\to\tkq$ is the canonical projection on the first factor, that is,
 $ (\tau^k_Q)^*\tau^k_Q(\mathbf{v}_{q},\mathbf{w}_{q})=\mathbf{v}_{q}\,,$

\bigskip

\noindent $\bullet$ \noindent   {\bf The map $\; {\bf i}: T^1_kQ \times_Q T^1_kQ
\longrightarrow   T(T^1_kQ)$.}

\bigskip

Let  us $V(T^1_kQ)=\left< \displaystyle\frac{\partial}{\partial v^i_A}\right>_{
1\leq i\leq n\,  ,\, 1\leq A \leq k}  $ denote  the vertical subbundle of $\taukq:\tkq\to
Q$ and   define the map
$$ {\bf i}: T^1_kQ \times_Q T^1_kQ \longrightarrow V(\tkq)\subset T(T^1_kQ)
$$
 by
$$
{\bf i}(\mathbf{v}_{q},\mathbf{w}_{q}) = \displaystyle\sum_{A=1}^k
\displaystyle\frac{d}{ds}\Big\vert_{0}({v_1}_{q},\ldots, {v_A}_{q
}+s{w_A}_{q},\ldots, {v_k}_{q})\,.  $$

This map is locally given by
\begin{equation}\label{locali}{\bf
i}(\mathbf{v}_{q},\mathbf{w}_{q})=\displaystyle\sum_{A=1}^k w^i_A
\displaystyle\frac{\partial}{\partial v^i_A}\Big\vert_{\mathbf{v}_{q}}\,.
\end{equation}

\bigskip

\noindent $\bullet$ \noindent{\bf Canonical vector fields on $\tkq$.}

\bigskip

  The canonical vector field (or Liouville vector field)  $\Delta\in \mathfrak{X}(\tkq)$ is defined by  $
\Delta(\mathbf{v}_q)={\bf i}(\mathbf{v}_q,\mathbf{v}_q)\,.$ This vector field is used to introduce the Energy Lagrangian function in the $k$-symplectic Lagrangian formalism, see section \ref{4.3}.

  From (\ref{locali})  we obtain that  its local expression is
\begin{equation}\label{locdelta}
\Delta =\displaystyle\sum_{A=1}^k\displaystyle\sum_{i=1}^{n} v^i_A
\displaystyle\frac{\partial}{\partial v^i_A} \quad .
\end{equation}

  The canonical vector fields $\Delta_A\in \mathfrak{X}(\tkq)$ are defined by
  $$\Delta_A(\mathbf{v}_q)={\bf i}(\mathbf{v}_q,(0,\ldots,\stackrel{A}{{v_A}_q}, \ldots , 0))\, $$  for all $ A=1, \ldots , k\,$, and they  
 are locally given by
\begin{equation}\label{locdeltaA}
\Delta_A= \sum_{i=1}^{n} v^i_A
\displaystyle\frac{\partial}{\partial v^i_A} \quad .
\end{equation}

\bigskip

\noindent $\bullet$ \noindent{\bf The vector bundle $(\tkq\times_Q
TQ,(\tau^k_Q)^*\tau_Q,\tkq)$.}

\bigskip

Let us consider now the fiber  $(\tau^k_Q)^*\tau_Q$, which is the
pull-back of the tangent bundle $TQ$ by $\tau^k_Q$.  This  fiber is also called
the {\bf transverse fiber to $\tau^k_Q$}. The total space of
  this  fiber is $$\tkq\times_Q TQ=
\{(\mathbf{v}_{q},u_{q} )\in \tkq\times TQ\;|
\;\taukq(\mathbf{v}_{q})=\tau_Q(u_{q}) \}\;.
$$ and
$(\taukq)^*\tau_Q:\tkq\times_Q TQ\to \tkq$ is the  canonical projection
 
 $$(\taukq)^*\tau_Q(\mathbf{v}_{q},u_{q})=\mathbf{v}_{q}\,.$$

\bigskip

\noindent $\bullet$ \noindent {\bf The map $\; {\bf j}: T(T^1_kQ)
\longrightarrow T^1_kQ \times_Q TQ$}

\bigskip

Let  $\tau_{\tkq}: T(\tkq) \to \tkq$ be  the tangent bundle of $\tkq$ and
$T\tau^k_Q:T(\tkq) \to TQ$ the tangent map of
$\tau^k_Q$. We define the map
$$\begin{array}{rcl}
{\bf j}:=(\tau_{\tkq},T\tau^k_Q) : T(T^1_kQ) & \to & T^1_kQ \times_Q TQ \\
\noalign{\medskip} Z_{\mathbf{v}_{q}} & \to &
 (\mathbf{v}_{q},T_{\mathbf{v}_{q}}\tau^k_Q(Z_{\mathbf{v}_{q}}))
\end{array}$$
 locally given by
\begin{equation}\label{localj}
{\bf j}(Z_{\mathbf{v}_{q}})={\bf j}\, \left(Z^i
\displaystyle\frac{\partial}{\partial q^i}\Big\vert_{\mathbf{v}_{q}}+ Z^i_A
\displaystyle\frac{\partial}{\partial v^i_A}\Big\vert_{\mathbf{v}_{q}}\right)=
\left(\mathbf{v}_{q},Z^i \displaystyle\frac{\partial}{\partial
q^i}\Big\vert_{q}\right) \, \, .
\end{equation}

The map ${\bf j}$ is a bundle homomorphism and the induce maps
  ${\bf j}_{\mathbf{v}_q}:T_{\mathbf{v}_q}(\tkq) \to \{\mathbf{v}_q\}\times T_qQ$
are linear, for all $\mathbf{v}_q\in\tkq$.

\begin{remark}    {\rm In Szilasi's book \cite{Szilasi}, page 62, one can be found the definition of   $\bf j$ for an arbitrary vector bundle $(E,\pi,M)$. In our case $E=\tkq$, $M=Q$ and
$\pi=\taukq$.}\qed\end{remark}

\noindent $\bullet$ {\bf The  short exact sequence arising from $\taukq$.}

\begin{lem} The sequence
\begin{equation}\label{sec}\xymatrix{0\ar[r] & \tkq\times_Q\tkq\ar[r]^-{\bf i}\ar[dr]
&T(\tkq)\ar[r]^-{\bf j}\ar[d] & \tkq\times_Q TQ\ar[r] \ar[dl]&0\\ & & T^1_kQ &&}\end{equation} is
a short exact sequence of vector bundles, that we will call the
{\bf canonical short exact  sequence} arising from $\tau^k_Q$.
\end{lem}
\begin{pf} This result can be proved for a general vector bundle $(E,\pi,M)$, see
 \cite{Szilasi}. In any case the principal point of the proof is that
 ${\bf j}\circ {\bf
i}=0 $,   which is a direct consequence of (\ref{locali}) and (\ref{localj}). \end{pf}

\noindent $\bullet${\bf Canonical $k$-tangent structure on $\tkq$}.
\bigskip

The canonical $k$-tangent structure is a certain family of $k$ tensor fields of type $(1,1)$. This structure was introduced by M. de Le\'{o}n {\it et al.} in \cite{LMS-1991}. Next we will describe an alternative definition of this structure.

We introduce the maps
${\bf k_A}$ from   $\tkq\times_Q TQ$ to
$\tkq\times_Q\tkq$ as follows
$$\begin{array}{cccl}
    {\bf k_A}: & T^1_kQ\times_Q TQ & \longrightarrow & T^1_kQ\times_Q T^1_kQ   \\
          & (\mathbf{v}_{q},u_{q})                &  \to &  (\mathbf{v}_{q},(0,\ldots,0,
          \stackrel{A}{u_{q}}, 0, \ldots, 0)
  \end{array}\quad1\leq A\leq k\,.
$$

The
composition $J^A=   {\bf i}\circ {\bf k_A} \circ {\bf j}\,\,$ is a tensor field on $\tkq$ of type $(1,1)$ locally given by
\[
{\small\xymatrix@R=0.5mm@C=5mm{T(T^1_kQ)\ar[r]^-{\bf j}\ar@/^{10mm}/[rrr]^-{J^A}& T^1_kQ\times_Q
TQ\ar[r]^-{\bf k_ A} & T^1_kQ\times_Q T^1_kQ \ar[r]^-{\bf i}&
T(T^1_kQ) \\
Z^i\derpar{}{q^i}\Big\vert_{\mathbf{v}_q}+Z^i_A\derpar{}{v^i_A}\Big\vert_{\mathbf{v}_q}\ar@{|->}[r]
&(\mathbf{v}_q,Z^i\derpar{}{q^i}\Big\vert_{q}
)\ar@{|->}[r]&(\mathbf{v}_q,(0,\ldots,\stackrel{A}{Z^i\derpar{}{q^i}\Big\vert_{q}},\ldots,
0))\ar@{|->}[r]&Z^i\derpar{}{v^i_A}\Big\vert_{\mathbf{v}_q}}}\]
or equivalently \begin{equation}\label{localJa}J^A= \derpar{}{v^i_A}\otimes dq^i \quad .\end{equation}
 The set
 $(J^1,\ldots,J^k)$ is called the  {\it  canonical $k$-tangent structure} on $\tkq$
, see   \cite{LMS-1991,MRS-2004,RSV-2007(2)}. Along this paper we will use this structure to characterize nonlinear connections on $\tau^k_Q\colon T^1_kQ\to Q$.

\section{Nonlinear connections on $\tau^k_Q:\tkq\to Q$.}\protect\label{Nc}

 Let us remember that an  {\it Ehresmann connection or nonlinear connection} on $\taukq:\tkq\to Q$ is a
 differentiable   subbundle    $H(\tkq)$ of $T(\tkq)$, called the
 horizontal subbundle of the connection,   which is complementary to the
   vertical subbundle $V(\tkq)$,  that is, $T(\tkq)=H(\tkq)\oplus V(\tkq).$

In this section we give two characterizations of the nonlinear connections on $\taukq:\tkq\to Q$. In the first one we use the canonical short exact sequence constructed in the above section in an analogous way to that one in Szilasi's book \cite{Szilasi} for the case $k=1$. After we characterize nonlinear connections on $\taukq:\tkq\to Q$ using the $k$-tangent structure $(J^1,\ldots, J^k)$. In the particular case $k=1$ this second result was obtained by Grifone \cite{grif1,grif2,grif3}.

\subsection{The horizontal maps.}\protect\label{Sec 4.2.1.}

\begin{definition}\label{ah}
  A right splitting of the short exact sequence
$$\xymatrix{0\ar[r] & \tkq\times_Q\tkq\ar[r]^-{\bf i}
&T(\tkq)\ar[r]^-{\bf j} & \tkq\times_Q TQ\ar[r] &0}\,,$$   is called a
{\it horizontal map  for   $\taukq$}. This map   is   a $\tkq$-morphism   of vector bundles (i.e. the morphism over the
base is $id_{\tkq}$)
$$ \mathcal{H}: T^1_kQ \times_Q TQ \longrightarrow
T(T^1_kQ) $$ satisfying
$${\bf j}\circ \mathcal{ H} =1_{ T^1_kQ \times_Q
TQ}\,.$$\end{definition}

Along this section we show that to give a horizontal map for $\tau^k_Q$ is equivalent to give a nonlinear connection on $\taukq:\tkq\to Q$.

\begin{prop}\label{h}
The horizontal map $ \mathcal{H}: T^1_kQ \times_Q TQ \longrightarrow
T(T^1_kQ)$ is locally given by
\begin{equation} \label{locH}  \mathcal{ H}(\mathbf{v}_{q},u_{q})=
u^i\left(\displaystyle\frac{\partial}{\partial q^i}\Big\vert_{ \mathbf{v}_{q}}-
 N^j_{Ai}(\mathbf{v}_{q})\displaystyle\frac{\partial}{\partial v^j_A}\Big\vert_{{\mathbf{v}_{q}}}\right)
\end{equation}
 where $\mathbf{v}_q\in\tkq,\,u_{q}\in TQ$ and the functions $N^j_{Ai}$,
defined on $\tkq$,  are called the components of the connection defined
 by $\mathcal{H}$.
\end{prop}
\begin{pf}  We write
$$ \mathcal{ H}(\mathbf{v}_{q},u_{q})=H^i(\mathbf{v}_{q},u_{q})
\displaystyle\frac{\partial}{\partial q^i}\Big\vert_{ \mathbf{v}_{q}}-
N^i_A(\mathbf{v}_{q},u_{q})\displaystyle\frac{\partial}{\partial
v^i_A}\Big\vert_{{\mathbf{v}_{q}}}
$$ for some functions on $H^i,\;N^i_A$ defined on $\tkq\times_Q TQ$.

 Since ${\bf j}\circ \mathcal{H}=1 _{T^1_kQ\times_Q
TQ}$, from (\ref{localj}), we obtain  \begin{equation}\label{Hlin}
\mathcal{ H}(\mathbf{v}_{q},u_{q})=u^i
\displaystyle\frac{\partial}{\partial q^i}\Big\vert_{ \mathbf{v}_{q}}-
N^i_A(\mathbf{v}_{q},u_{q})\displaystyle\frac{\partial}{\partial
v^i_A}\Big\vert_{{\mathbf{v}_{q}}}\,.\end{equation}

On the other hand, the induced maps
$$\mathcal{H}_{\mathbf{v}_q}:(\tkq\times_Q TQ)_{\mathbf{v}_q}\cong\{\mathbf{v}_q\}\times
T_{q}Q\to T_{\mathbf{v}_q}(\tkq)$$ are linear for all  $\mathbf{v}_q\in\tkq$, then from (\ref{Hlin})
we obtain  that
\begin{equation} \label{loccalh}\begin{array}{lcl}  \mathcal{ H}(\mathbf{v}_{q},u_{q}) &=&
  \mathcal{ H}\left(\mathbf{v}_{q},u^i\derpar{}{q^i}
\Big\vert_{q}\right)= u^i\mathcal{ H}\left(\mathbf{v}_{q},\displaystyle\frac{\partial}{\partial q^i}\Big\vert_{{\mathbf{v}_{q}}}
\right) \\\noalign{\medskip} & =&
u^i\left(\displaystyle\frac{\partial}{\partial q^i}\Big\vert_{ \mathbf{v}_{q}}-
N^j_A(\mathbf{v}_{q}, \displaystyle\frac{\partial}{\partial
q^i}\Big\vert_{q} )\displaystyle\frac{\partial}{\partial
v^j_A}\Big\vert_{{\mathbf{v}_{q}}}\right)\,.  \end{array}\end{equation}

Now   defining the functions $N^j_{Ai}$ on $\tkq$ by
$$ N^j_{Ai}(\mathbf{v}_{q})=N^j_A(\mathbf{v}_{q}, \displaystyle\frac{\partial}{\partial
q^i}\Big\vert_{q} ),\qquad  1\leq i,j\leq n,\, 1\leq A\leq k$$
  we obtain (\ref{locH}).\end{pf}

To each horizontal map $\mathcal{H}:\tkq\times_Q TQ\to
T(\tkq)$ we associate  the horizontal and vertical projectors as follows:
\begin{enumerate}
\item {\it The horizontal projector} is given by
 ${\bf h}:\,=\mathcal{ H}\circ {\bf j}:T(\tkq) \to T(\tkq)$.

From (\ref{localj})
we deduce that the local expression of ${\bf h}$ is 
\begin{equation}\label{bfh} {\bf  h}= \left(\displaystyle\frac{\partial}{\partial q^i} -
N^j_{Ai}\displaystyle\frac{\partial}{\partial v^j_A}\right)\otimes dq^i
\end{equation} and satisfies   ${\bf h}^2={\bf h}, \;   Ker\, {\bf
h}=V(\tkq) $
and
$$Im\, {\bf  h} =\left<\displaystyle\frac{\partial}{\partial q^i} -
N^j_{Ai}\displaystyle\frac{\partial}{\partial v^j_A}\right>_{i=1,\ldots , n}
\quad . $$

\item {\it The vertical projector}  is given by  ${\bf v}:\,=1_{T(\tkq)}-{\bf h}$ and it satisfies
  $${\bf v}^2={\bf v}, \quad , \quad Ker\, {\bf
v}=Im\,{\bf h}  \quad  , \quad Im\, {\bf v} = V(\tkq)$$

From (\ref{bfh})  we obtain
\begin{equation}\label{localv}
 {\bf v}=\displaystyle\frac{\partial}{\partial v^j_A}\otimes(
dv^j_A+N^j_{Ai} dq^i)\;.
\end{equation}

 Since  ${\bf v}:\,=1_{T( T^1_kQ )}-{\bf h}$ and ${\bf h}^2={\bf h}$ we obtain that ${\bf v}{\bf h}={\bf h}{\bf v}=0$.
\end{enumerate}

The following Lemma is well known
\begin{lem}\label{new} Let $M$ be an arbitrary manifold and $\Gamma$ an almost product structure, i.e., $\Gamma$ is a tensor field of type $(1,1)$ such that $\Gamma^2= 1_M$. If we put
$$
{\bf h}=\displaystyle\frac{1}{2} (1_M+\Gamma) \quad , \quad {\bf v}=   \displaystyle\frac{1}{2} (1_M-\Gamma)
$$ then
\begin{equation}\label{ch}{\bf h}^2={\bf h} \quad {\bf h}{\bf v}={\bf v}{\bf h}=0 \quad {\bf v}^2={\bf v}\quad  .
\quad
\end{equation}
Conversely if $\,\, {\bf h}\, $ and $\, {\bf v}\, $ are two tensor fields of type $(1,1)$ and they satisfy (\ref{ch}) then $\, \Gamma={\bf h} -{\bf v}\, $ is an almost product structure, and   we have  
$TM=Im {\bf h}\oplus Im {\bf v}$.
\end{lem}
Then, in our case $M=\tkq$  we have $$T(\tkq)=Im\, {\bf h}\oplus Im\, {\bf v}= Im\, {\bf h}\oplus V(\tkq)$$
Thus $Im\, {\bf h}$ is the nonlinear connection associated to $\mathcal{H}$.

We have seen that each horizontal map $\mathcal{H}$ corresponds  an horizontal projector ${\bf h}$ which defines a nonlinear connection on $T^1_kQ$. The converse of this is given in   Lema 1, page 72 Szilasi \cite{Szilasi}, for an arbitrary vector bundle; in our case one obtains
 \begin{lem}\label{htoH}
If ${\bf h}\in \mathcal{T}^1_1(\tkq)$ is an horizontal projector
for $\tau^k_Q$,  that is   ${\bf h}^2={\bf h}\quad \makebox{and} \quad Ker\;{\bf
h}=V(\tkq)$,  then there exists an unique  horizontal map $
\mathcal{H}: T^1_kQ \times_Q TQ \longrightarrow T(T^1_kQ) $ such that $\mathcal{H}\circ{\bf j}={\bf h}$.
 \end{lem}
 \qed

Let  $X\in \mathfrak{X}(Q)$ be a  vector field on $Q$. Then the {\it
 horizontal lift} $X^h$ of $X$ to $\mathfrak{X}(\tkq)$ is defined as follows
  \begin{equation}\label{horil} X^h(\mathbf{v}_{q}):\,=\mathcal{H}(\mathbf{v}_{q},X(q))
  =X^i\left(\displaystyle\frac{\partial}{\partial q^i}\Big\vert_{\mathbf{v}_{q}}-
  N^j_{Ai}(\mathbf{v}_{q})\displaystyle\frac{\partial}{\partial
  v^j_A}\Big\vert_{\mathbf{v}_{q}}\right)\; ,
\end{equation} being $X=X^1\frac{\partial}{\partial q^i}$.

The {\it curvature } $\Omega:\mathfrak{X}(\tkq)\times \mathfrak{X}(\tkq) \to
\mathfrak{X}(\tkq)$ of the horizontal map $\mathcal{H}$ is defined as $\Omega=-\displaystyle\frac{1}{2}[\mathbf{h},\mathbf{h}]$
and it is locally given by
\begin{equation}\label{loccur}\Omega=\displaystyle\frac{1}{2}\left(\derpar{N^j_{Ak}}{q^i} -
\derpar{N^j_{Ai}}{q^k} + N^m_{Bk}\derpar{N^j_{Ai}}{v^m_B}
-N^m_{Bi}\derpar{N^j_{Ak}}{v^m_B} \right)\derpar{}{v^j_A}\otimes
dq^i\wedge dq^k\,.\end{equation}

\subsection{Nonlinear connections and canonical $k$-tangent structure on
$\tkq$.}\protect\label{3.2}

In this section we characterize  nonlinear connections on $\tkq$ using the  canonical $k$-tangent
structure $(J^1, \ldots, J^k)$.

\begin{prop}\label{new1}Let  $\Gamma$ be a   tensor field of type $(1,1)$ on
$\tkq$ satisfying \begin{equation}
\label{conex} J^A\circ  \Gamma =  J^A \quad \makebox{and}
\quad
 \Gamma \circ J^A  =    -J^A \,,\;1\leq A\leq k\,.
\end{equation} Then, $\Gamma$ is an almost product structure, that is ,
 $\Gamma^2= 1_{\tkq}$.
\end{prop}
\begin{pf}  For each vector field $Z$ on $\tkq$ we
have $J^A(\Gamma Z)=  J^A(Z),\; 1\leq A\leq k,$ then  $J^A(\Gamma
(Z)-Z)=0,$ that is, the vector field $\Gamma (Z)- Z$ is
vertical, then it can be written as follows:$$\Gamma
(Z)-Z=\displaystyle\sum_{B=1}^kJ^B(W_B),$$ where $W_1, \ldots, W_k$ are vector fields on  $T^1_kQ$.  Finally we obtain
$$\begin{array}{lcl}\Gamma^2(Z)&=& \Gamma( \Gamma (Z))=\Gamma
(Z+\displaystyle\sum_{B=1}^kJ^B(W_B))=  \Gamma ( Z)+ \displaystyle\sum_{B=1}^k  \Gamma
(J^B(W_B)) \\\noalign{\medskip} &=& \Gamma ( Z) -\displaystyle\sum_{B=1}^k
J^B(W_B)=Z \quad .\end{array}$$\end{pf}

From (\ref{conex}) we deduce that
   $\Gamma$ is locally given by
\begin{equation}\label{Gamma}
 \Gamma  = \left( \frac{\partial}{\partial q^i} + \Gamma^j_{Ai}
\frac{\partial}{\partial v^j_A}\right)  \otimes dq^i -
\frac{\partial}{\partial v^i_A} \otimes dv^i_A\;,
\end{equation} where $\Gamma^j_{Ai}$ are functions defined on
$\tkq$ called {\it the components of $\Gamma$}.

\begin{prop} To give a nonlinear connection $N$ on $\taukq:\tkq\to Q$ is equivalent to give a tensor field $\Gamma$ of type $(1,1)$ satisfying (\ref{conex}).\end{prop}

\begin{pf} Let $N$ be  a nonlinear connection on $\taukq:\tkq\to Q$ with horizontal projector $\mathbf{h}$. Then $\Gamma=2\mathbf{h}-1_{\tkq}$ satisfies (\ref{conex}). In fact, one obtains:
$$J^A\circ \Gamma= 2(J^A\circ \mathbf{h})- J^A=2J^A-J^A=J^A$$ where we have used that $J^A\circ \mathbf{h}= J^A$.
On the other hand, since $\mathbf{h}\circ J^A=0$ we  
$$\Gamma\circ J^A=2(\mathbf{h}\circ J^A)-J^A=-J^A\,.$$

Conversely, given $\Gamma$ satisfying (\ref{conex}) from the above proposition we obtain that $\Gamma^2=1_{\tkq}$, then from   Lemma \ref{new} we deduce that there exists a horizontal projector
${\bf h}=\displaystyle\frac{1}{2}(1_{T(T^1_kQ)}+\Gamma)$, with local expression
$$
{\bf h}=\displaystyle\frac{1}{2}(1_{T(T^1_kQ)}+\Gamma) = \left( \frac{\partial}{\partial q^i} + \displaystyle\frac{1}{2} \Gamma^j_{Ai}
 \frac{\partial}{\partial v^j_A}\right)  \otimes dq^i  \;,
$$
which defines a nonlinear connection  $N_\Gamma$. Moreover
 the components of the nonlinear connection $N_\Gamma$   are given by
  $$(N_\Gamma)^j_{Ai}= -\displaystyle\frac{1}{2}\Gamma^j_{Ai} \,.$$\end{pf}

\section{$k$-vector fields. Second order partial differential equations ({\sc sopde}s).}
\label{sopde}

Second order differential equations, usually called {\sc sodes}  play an important role on the geometric description of Lagrangian Mechanics.

In this section we introduce  {\sc sopde}s (second order partial differential equations) which are a generalization of the concept of {\sc sode}. We study the relationship between {\sc sopde}s and nonlinear connections on $T^1_kQ$ and we also show the role of {\sc sopde}s in Lagrangian classical field theories. Let us observe that the role of {\sc sopde's} in the
  $k$-symplectic \cite{GU-1987,MRS-2004,RSV-2007(2)} and $k$-cosymplectic
 \cite{LMORS-1998} Lagrangian formalisms of classical field theories is very important and similar to role of
second-order differential equations, {\sc sode}'s, in Lagrangian Mechanics.

\begin{definition}  \label{kvector} Let $M$ be an arbitrary manifold
and $\tau_M^k : T^{1}_{k}M \longrightarrow M$ its  tangent bundle of
$k^1$-velocities.
A {\it $k$-vector field} on $M$ is a section ${\bf \xi} : M \longrightarrow T^1_kM$ of the projection
$\tau_M^k$.
\end{definition}

Since $T^{1}_{k}M$ is  the Whitney sum $TM\oplus \stackrel{k}{\dots}
\oplus TM$ of $k$ copies of $TM$,
  we deduce that a $k$-vector field ${\bf \xi}$ defines
a family of $k$ vector fields $\{\xi_{1}, \dots, \xi_{k}\}$ on $M$ by
projecting  ${\bf \xi}$ onto every factor. For this reason we will
denote a $k$-vector field ${\bf \xi}$ by $(\xi_1, \ldots, \xi_k)$.

\begin{definition} \label{integsect}
An {\it integral section}  of a $k$-vector field  \, ${\bf
\xi}=(\xi_{1}, \dots, \xi_{k})$ \, passing through a point $x\in M$ is a
map
  $\phi:U_\mathbf{0}\subset \r^k \rightarrow M$, defined on some neighborhood  $U_\mathbf{0}$ of $\mathbf{0}\in \rk$,
  such that
\begin{equation}\label{anflo}
\phi(\mathbf{0})=x \quad , \quad \phi_*(\mathbf{t})\left(
\displaystyle\frac{\displaystyle\partial}{\displaystyle\partial
t^A}\Big\vert_\mathbf{t} \right) = \xi_{A}(\phi (\mathbf{t})) \, \quad \mbox{for every}
\quad \mathbf{t}\in U_\mathbf{0}, \quad ,
\end{equation}
 or equivalently,    $\phi$ satisfies
$ {\bf \xi} \circ\phi=\phi^{(1)}, $ where  $\phi^{(1)}$ is the first
prolongation of $\phi$ defined by
$$
\begin{array}{rccl}\label{1prolong}
\phi^{(1)}: & U_\mathbf{0}\subset \r^k & \longrightarrow & T^1_kM \\
\medskip
 & \mathbf{t} & \longrightarrow & \phi^{(1)}(\mathbf{t})=j^1_\mathbf{0}\phi_\mathbf{t} \quad ,
 \quad \phi_{\mathbf{t}} (\mathbf{t})=\phi (\bar{\mathbf{t}}+\mathbf{t}) \end{array},
$$
for every $ \mathbf{t},\mathbf{\bar{t}} \in \r^k$ such that $ \mathbf{\bar{t}}+\mathbf{t}\in U_\mathbf{0}$.

A $k$-vector field ${\bf \xi}=(\xi_1,\ldots , \xi_k)$ on $M$
is said to be {\it integrable} if there is an integral section passing through each
point of $M$.
\end{definition}

In local coordinates one obtains
\begin{equation}\label{localfi11}
\phi^{(1)}(t^1, \dots, t^k)=\left( \phi^i (t^1, \dots, t^k),
\displaystyle\frac{\displaystyle\partial\phi^i}{\displaystyle\partial
t^A} (t^1, \dots, t^k)\right) \quad  .
\end{equation}

\begin{remark}Let us observe that in the case $k=1$, an integral section is an integral curve and the first prolongation is the tangent lift from a curve on $M$ to $TM$.\end{remark}

Next we will introduce the notion of {\sc sopde}, which is
 a class of $k$-vector fields on $\tkq$.

We shall see that the integral sections of {\sc sopde}s are
first prolongations $\phi^{(1)}$ of maps $\phi:\rk\to Q$.

If $F:M\to N$ is a differentiable map  between the manifolds $M$ and $N$, then $T^1_kF:T^1_kM \to    T^1_kN$ is defined by $T^1_kF({v_1}_q, \ldots, {v_k}_q)=
(F_*(q)({v_1}_q), \ldots , F_*(q){v_k}_q)$, or equivalently $T^1_kF(j^1_0\sigma)=j^1_0(F\circ \sigma)$.

\begin{definition} \label{sode0}
A $k$-vector field ${\bf \xi}=(\xi_1 ,\ldots,\xi_k) $ on $T^1_kQ$ is  a {\it second
order partial differential equation ({\sc sopde})} if it is also a
section of the projection
$T^1_k(\tau^k_Q):T^1_k(T^1_kQ)\rightarrow T^1_kQ$; that is,
\begin{equation}
T^1_k(\tau^k_Q)\circ{\bf \xi}=1_{T^1_kQ}\,. \label{sodedef}
\end{equation}
\end{definition}

Let us observe that $\xi_A\in \mathfrak{X}(\tkq)$ and (\ref{sodedef}) means
$$(\tau^k_Q)_*(\mathbf{v}_q)(\xi_A(\mathbf{v}_q))={v_A}_q \quad A=1, \ldots, k \, .$$
where $\mathbf{v}_q=({v_1}_q, \ldots,{v_k}_q)$.

Let $(q^i)$ be a local coordinate system on $U\subseteq Q$ and $(q^i,v^i_A)$ the
induced local coordinate system on $T^1_kU$. From (\ref{sodedef}), a
direct computation   shows that the local
expression of a {\sc sopde} ${\bf \xi}=(\xi_1 ,\ldots,\xi_k) $ is
\begin{equation}\label{localsode1}
\xi_A(q^i,v^i_A)= v^i_A\frac{\displaystyle
\partial} {\displaystyle
\partial q^i}+
(\xi_A)^i_B \frac{\displaystyle\partial} {\displaystyle
\partial v^i_B},\quad 1\leq A \leq k \quad .
\end{equation}
where $(\xi_A)^i_B\in\mathcal{C}^\infty(T^1_kU)$.

 If $\varphi:\rk \to
  T^1_kQ$,   is an integral section of a {\sc sopde} $(\xi_1
,\ldots,\xi_k) $ locally given by
$\displaystyle\varphi(\mathbf{t})=\left(\varphi^i(\mathbf{t}),\varphi^i_B(\mathbf{t})\right)$ then
$\xi_A(\varphi(\mathbf{t}))=\varphi_*(\mathbf{t})[\partial /\partial t^A(\mathbf{t})]$ and thus
\begin{equation} \label{fit}\frac{\displaystyle\partial\varphi^i}
{\displaystyle\partial t^A}(\mathbf{t})=v^i_A(\varphi(\mathbf{t}))=\varphi^i_A(\mathbf{t})\,
,\qquad \frac{\displaystyle\partial\varphi^i_B}
{\displaystyle\partial t^A}(\mathbf{t})=(\xi_A)^i_B(\varphi(\mathbf{t}))\, .
\end{equation}

 From (\ref{localfi11}) and (\ref{fit}) we obtain:

\begin{prop} \label{sope1}
Let ${\bf \xi}=(\xi_1 ,\ldots,\xi_k) $ be an integrable {\sc sopde} on
$\tkq$. If $\varphi$ is an integral section of $\xi$ then
$\varphi=\phi^{(1)}$, where $\phi^{(1)}$ is the first prolongation of
the map
$\phi=\tau^k_Q\circ\varphi:\rk\stackrel{\varphi}{\to}T^1_kQ\stackrel{\tau^k_Q}{\to}Q$
and it is a solution to the system
\begin{equation}\label{nn1}
 \frac{\displaystyle\partial^2 \phi^i} {\displaystyle\partial t^A
\partial t^B       }(\mathbf{t})= (\xi_A)^i_B(\phi^{(1)}(\mathbf{t})) =(\xi_A)^i_B(\phi^i(\mathbf{t}),
\displaystyle\frac{\partial\phi^i}{\partial t^C}(\mathbf{t}))\,.
\end{equation}
Conversely, if $\phi:\rk \to Q$ is any map satisfying  (\ref{nn1}),
then $\phi^{(1)}$ is an integral section of ${\bf \xi}=(\xi_1
,\ldots,\xi_k) $.
\end{prop}

\begin{remark}\label{ig}
{\rm For an integrable {\sc sopde} we have $(\xi_A)^i_B=(\xi_B)^i_A$}.
\end{remark}
The following characterization of {\sc sopde}s can be given using
the canonical $k$-tangent structure of $T^1_kQ$ (see
 (\ref{locdeltaA}), (\ref{localJa}) and (\ref{localsode1})):

\begin{prop}
    \label{pr235}
A $k$-vector field $\mathbf{\xi}=(\xi_1,\ldots,\xi_k)$ on
$T^1_kQ$ is a {\sc sopde}  if, and only if,
$S^A(\Gamma_A)=\Delta_A$, for all $A= 1, \ldots , k$\ .
\end{prop}
\begin{example}
Let us consider the following {\sc sopde} $(\xi_1,\xi_2)$ on $T^1_2\r$ , with coordinates $(q,v_1,v_2)$, given by
\begin{eqnarray} \label{gamheat}
\begin{array}{lcl}
\xi_1 & = &v_1 \displaystyle\frac{\partial }{\partial q}
-\displaystyle\frac{k}{\lambda^2}\, v_1 \,  \frac{\partial }{\partial v_1} +-\displaystyle\frac{k}{\lambda^2}\, v_2 \, \frac{\partial }{\partial v_2}
  \\
\noalign{\medskip} \xi_2 & = & v_2 \displaystyle\frac{\partial}{\partial q}+
-\displaystyle\frac{k}{\lambda^2}\, v_2 \, \displaystyle\frac{\partial }{\partial v_1}+
+\displaystyle\frac{1}{k}\, v_1 \,
\frac{\partial}{\partial v_2}
\end{array}
\end{eqnarray}
Let $\phi:(t,x)\in\r^2 \to \r$ be a map. If $\phi^{(1)}\colon\r^2\to T^1_ 2\r$ is an integral section of $(\xi_1,\xi_2)$ then
from (\ref{nn1}) we obtain
\begin{eqnarray}
 -\displaystyle\frac{k}{\lambda^2}  \displaystyle\frac{\partial \phi}{\partial t} &
 =& \displaystyle\frac{\partial^2 \phi}{\partial t^2} \label{o1}
 \\ \noalign{\medskip}
 -\displaystyle\frac{k}{\lambda^2}  \displaystyle\frac{\partial \phi}{\partial x} &
 =& \displaystyle\frac{\partial^2 \phi}{\partial t \partial x}\label{o2}
 \\ \noalign{\medskip}
 \displaystyle\frac{1}{k} \displaystyle\frac{\partial \phi}{\partial t} &
 =& \displaystyle\frac{\partial^2 \phi}{\partial x^2} \label{eccalor}
\end{eqnarray}
Equation (\ref{eccalor}) is the one-dimensional  heat equation   where $k$ is the thermal diffusivity  and the solutions  $\phi(t,x)$ represents the temperature at the point x of a rod at time $t$ .

Any  integral section of this {\sc sopde} is the first prolongation of a solution of the heat equation. The general solution of (\ref{eccalor}) is
$$
\phi(t,x)= e^{-\frac{\kappa }{\lambda^2}t}\left[C
\cos\left(\frac{x}{\lambda}\right) +D
  \sin\left(\frac{x}{\lambda}\right)\right]=A\,e^{-\frac{\kappa }{\lambda^2}t}\sin
  \left(\frac{x}{\lambda}+\delta\right)
$$
 where $\lambda$, $C$ and $D$ are arbitrary constants
  and $A=\sqrt{C^2+D^2 }$, $\tan\delta =\displaystyle\frac{C}{D}$. Thus any solution of (\ref{eccalor}) is solution of  (\ref{o1}) and (\ref{o2}).\end{example}

\subsection{Relationship between {\sc sopde}s and  nonlinear connections .}\protect\label{4.2}

In this section we prove that each nonlinear connection defines a second order partial differential equation ({\sc sopde}) on $\tkq$ and conversely,  given a {\sc sopde} $\xi$ on $\tkq$   a nonlinear connection $N_\xi$ on $\tau^k_Q\colon \tkq\to Q$ can be defined.

\bigskip

\noindent $\bullet$ {\bf {{\sc sopde} associated to a nonlinear connection
 .}}\label{scnl}

\bigskip
Let us consider a nonlinear connection
   on $\taukq:\tkq\to Q$ with horizontal map
$\mathcal{H}:\tkq\times_Q TQ\to T(\tkq)$.
  For each $A= 1,\ldots ,  k$ we define $\xi_{\mathcal{H}}^A\in \mathfrak{X}(\tkq)$ as follows $$\xi_{\mathcal{H}}^A(\mathbf{v}_{q})=
  \mathcal{H}(\mathbf{v}_{q},v_{A_{\bf q}})\qquad\makebox{where } \quad\mathbf{v}_{\bf q}=({v_1}_q ,\ldots , {v_k}_q)\in \tkq$$

 From
(\ref{locH})   we obtain
that the {\sc sopde} $\xi_{\mathcal{H}}=(\xi_{\mathcal{H}}^1,
\ldots, \xi_{\mathcal{H}}^k)$ associated to $\mathcal{H}$ is
\begin{equation}\label{xicalh}
\xi_{\mathcal{H}}^A(\mathbf{v}_{q}) =v^i_A
\left(\displaystyle\frac{\partial}{\partial q^i}\Big\vert_{{\mathbf{v}_{q}}}-N^k_{Bi} \displaystyle\frac{\partial}{\partial v^k_B}\Big\vert_{
{\mathbf{v}_{q}}}\right) \quad .\end{equation}

\bigskip

\noindent $\bullet$ {\bf   Nonlinear connection  associated to a  {\sc sopde}.}

\begin{thm}\label{thxi}
To each {\sc sopde} $\xi$ on $\tkq$ we associate a nonlinear connection  $N_\xi$ with horizontal projector
\begin{equation}\label{hxi}
{\bf h}_{\xi}= \displaystyle\frac{1}{k+1}\, (1_{\tkq} - \, \displaystyle\sum_{A=1}^k\mathcal{
L}_{\xi_A}J^A)\,.
\end{equation}
\end{thm}

\begin{pf} Let  $\xi=(\xi_1,\ldots, \xi_k)$ be a {\sc sopde} on $\tkq$
locally given by
 $$\xi_A=v_A^i\derpar{}{q^i} +(\xi_{A})_B^j\derpar{}{v^j_B} \;,\quad A=1,\ldots, k\, .$$

Since
 $\mathcal{L}_{\xi_A}J^A(Z)=[\xi_A,J^AZ]-J^A[\xi_A,Z]$  for   all
vector field $Z$ on $\tkq$, we obtain
$$ \displaystyle\sum_{A=1}^k\mathcal{L}_{\xi_A}S^A = -\left(k\frac{\partial}{\partial
q^i} + \sum_{A=1}^k\frac{\partial (\xi_A)^j_B}{\partial v^i_A}
\frac{\partial }{\partial v^j_B}\right) \otimes dq^i +
\frac{\partial }{\partial v^i_B}\otimes dv^i_B$$
then a straightforward computation in local coordinates shows that   ${\bf h}_{\xi}$ is locally given by
\begin{equation}\label{local nabla xi}
{\bf h}_\xi =  \left( \displaystyle\frac{\partial}{\partial q^j}
 +\displaystyle\frac{1} {k+1}\displaystyle\sum_{A=1}^k
  \displaystyle\frac{ \partial (\xi_A)^i_B}{ \partial v^j_A}
\displaystyle\frac{\partial}{\partial v^i_B} \right) \otimes dq^j\,.
\end{equation}
and  satisfies
$${\bf h}_{\xi}^2={\bf h}_{\xi}\qquad \makebox{and} \qquad Ker\,{\bf
h}_{\xi}=V(\tkq)
$$
so defining $\mathbf{v}_\xi=1_{\tkq}- {\bf h}_{\xi} $ we obtain, see Lemma \ref{new}, that
$T(\tkq)=Im\, {\bf h}_{\xi}\oplus V(\tkq)$.
 \end{pf}

\begin{remark}
In the case $k=1$, the horizontal projector ${\bf h}_\xi$ given in
(\ref{hxi}), coincides with the projector given  by Grifone
\cite{grif1,grif3} and by Szilasi \cite{Szilasi}. \end{remark}

\begin{remark} \

 \begin{enumerate}
\item
 From  (\ref{bfh}) and  (\ref{local nabla xi}) we deduce that the components of the connection $N_\xi$
 are given by
\begin{equation}\label{compgamxi}
 (N_\xi)^\imath_{Bj} =- \displaystyle\frac{1}{k+1}\displaystyle\sum_{A=1}^k
  \displaystyle\frac{ \displaystyle\partial (\xi_A)^\imath_{B}}{ \displaystyle\partial v^j_A} \quad .
\end{equation}
\item We can associate to each {\sc sopde} $\xi$ the  almost product structure $\Gamma_\xi= 2{\bf h}_\xi-1_{\tkq}$, locally  given by
$$
\Gamma_\xi=  \displaystyle\frac{1}{k+1}\left((1-k)1_{\tkq} - 2\displaystyle\sum_{A=1}^k \,
\mathcal{L}_{\xi_A}J^A)\right)
$$
 In the case  $k=1$,  this  tensor field is
$\Gamma_\xi= -\mathcal{L}_\xi J$, where $J$ is the canonical tangent structure on $TQ$. The nonlinear connection associated to this structure was introduced by Grifone in   Proposition
I.41 of \cite{grif1} and   Proposition 1.3 of \cite{grif3}.
\end{enumerate}
\end{remark}

Along this section we have showed that there is a correspondence such that to each nonlinear connection on $\tkq$ we can associate a {\sc sopde} $\xi$ and conversely, given a {\sc sopde} on $\tkq$ there exists a nonlinear connection associated to this {\sc sopde}. Is this correspondence a bijection? In general the answer to this question is negative. In fact:

\begin{enumerate}
\item Let $\xi$ be a {\sc sopde} and $N_{\xi}$ be the nonlinear connection associated to $\xi$. We denote by $H_{\xi}$ the horizontal map associated to $N_\xi$. From (\ref{xicalh}) and (\ref{compgamxi})   we deduce  that  $\xi=\xi_{\mathcal{H}_\xi} $ if and only if
 $$
(\xi_A)_B^j=\displaystyle\frac{1}{k+1}\displaystyle\sum_{C=1}^k \displaystyle\frac{ \displaystyle\partial
(\xi_C)^j_B}{ \displaystyle\partial v^i_C}\, v^i_A \,,1\leq A,B\leq k,\,1\leq
i\leq n\, .
$$

When  $k=1$ we obtain  $\xi_{\mathcal{H}_\xi}=\xi$
if and only if
$
   \displaystyle\frac{1}{2}\displaystyle\frac{ \displaystyle\partial \xi^k}{
\displaystyle\partial v^i}\, v^i=\xi^k
$
 which means that the functions $\xi^k$ are homogeneous of grade $2$ (see \cite{grif3}).
\item Let us consider now a nonlinear connection   $N$ defined from a
  horizontal map $\mathcal{H}$, the {\sc sopde}
$\xi_{\mathcal{H}}$ associated to this connection and the  connection
$N_{\xi_{\mathcal{H}}}$ associated to the  {\sc sopde} $\xi_{\mathcal{H}}$.

From (\ref{xicalh})
 and (\ref{compgamxi}) we obtain  that
  $N=N_{\xi_{\mathcal{H}}}$ if and only if
$$N^j_{B\,i}=
v^l_A\derpar{N^j_{Bl}}{v^i_A},\qquad \,1\leq i,j\leq n,\, 1\leq B\leq
k\,.$$
\end{enumerate}

\subsection{{\sc sopde}s on Classical Field Theory}\label{4.3}

    In this subsection, we
recall  the Lagrangian formalism developed by G\"unther \cite{GU-1987}, see also \cite{MRS-2004}. Here we show  the role of sopdes and its integral sections in the Lagrangian Field Theory.

  Let $L\colon T^1_kQ  \to \mathbb{R}$ be a Lagrangian, that is a function $L(\phi^i,\partial \phi^i/\partial t^A)$ that depend on the components of the field and on its first partial derivatives. This Lagrangian is called autonomous in the sense that not depends on the time-space variables $(t^A)$ .

The {\sl generalized Euler-Lagrange equations} for $L$ are:
\begin{equation}
\label{ELe} \sum_{A=1}^k\frac{\partial}{\partial t^A}\Big\vert_t
\left(\frac{\displaystyle\partial L}{\partial
v^i_A}\Big\vert_{\psi(\mathbf{t})}\right)= \frac{\partial L}{\partial
q^i}\Big\vert_{\psi(\mathbf{t})}
 \quad , \quad
v^i_A(\psi(\mathbf{t}))= \frac{\partial\psi^i}{\partial t^A}
\end{equation}
whose solutions are maps $\psi\colon\mathbb{R}^k \to T^1_kQ$ with $\psi(\mathbf{t})=(\psi^i(\mathbf{t}),\psi^i_A(\mathbf{t}))$. Let us
observe that $\psi(\mathbf{t})=\phi^{(1)}(\mathbf{t})$, for  $\phi=\tau_Q^k \circ
\psi$.  Using the canonical $k$-tangent structure, one introduces a family of $1$-forms  $\theta_L^A$ on $T^1_kQ$, and a family of $2$-forms  $\omega_L^A$ on $T^1_kQ$ , as follows
\begin{equation}\label{betaloc}
 \theta_L^A=  dL \circ J^A  \quad , \quad  \omega_L^A=-\d\theta_L^A \quad , \quad 1 \leq
A \leq k  \quad .
\end{equation}

In local natural coordinates we have \begin{eqnarray} \label{thetala}
\theta_L^A = \displaystyle\frac{\partial L}{\partial v^i_A}\, dq^i\quad ,\quad
\label{omegala} \omega_L^A =  \frac{\partial
^2 L}{\partial q^j\partial v^i_A}\d q^i\wedge\d q^j +
\frac{\partial ^2 L}{\partial v^j_B\partial v^i_A}\d q^i\wedge\d
v^j_B \ . \end{eqnarray}

We also introduce the {\sl Energy  function}
$E_L=\Delta(L)-L\in\Cinfty(T^1_kQ)$, whose local expression is
\begin{equation}E_L=v^i_A\frac{\partial L}{\partial v^i_A}-L \ .
\label{energyL} \end{equation}

\begin{definition}
 The Lagrangian $L\colon T^1_kQ\longrightarrow \mathbb{R}$ is
said to be {\it regular} if the matrix
 $\left(\frac{\partial^2 L}{\partial v^i_A \partial v^j_B}\right)$ is not singular
at every point of $T^1_kQ$.
\end{definition}

Let  $ (\xi_1,\dots,\xi_k)$ be  a $k$-vector field on
$T^1_kQ$  locally given by
$$
\xi _A  =  ( \xi _A)^i \frac{\partial}{\partial  q^i} + (
\xi _A)^i_B\frac{\partial}{\partial v^i_B}\quad .
$$
Then from (\ref{thetala}) and  (\ref{energyL}) we deduce that  $ (\xi_1,\dots,\xi_k)$
 is a solution  to the equation
 \begin{equation}
\label{genericEL} \sum_{A=1}^k \inn_{\xi_A}\omega_L^A=\d E_L\, .
 \end{equation}
 if, and only if, $( \xi_A)^i$ and $( \xi _A)^i_B$ satisfy the system of equations
\begin{eqnarray*}
  \left( \frac{\partial^2 L}{\partial q^i \partial v^j_A} - \frac{\partial^2 L}{\partial q^j \partial v^i_A}
\right) \, ( \xi _A)^j - \frac{\partial^2 L}{\partial v_A^i
\partial v^j_B} \, ( \xi _A)^j_B &=&
 v_A^j \frac{\partial^2 L}{\partial q^i\partial v^j_A} - \frac{\partial  L}{\partial q^i } \, ,
\\
\frac{\partial^2 L}{\partial v^j_B\partial v^i_A} \, ( \xi_A)^i
 &=& \frac{\partial^2 L}{\partial v^j_B\partial v^i_A} \, v_A^i \quad .
\end{eqnarray*}

If the Lagrangian is regular, the above equations  are equivalent
to the equations \begin{eqnarray} \label{locel4} \frac{\partial^2 L}{\partial
q^j \partial v^i_A} v^j_A + \frac{\partial^2 L}{\partial
v_A^i\partial v^j_B}( \xi _A)^j_B = \frac{\partial  L}{\partial
q^i}
\\
\label{locel3}
 ( \xi _A)^i= v_A^i\quad , \quad 1\leq i \leq n, \,  1\leq A \leq k\ .
\end{eqnarray}

Thus, if $L$ is a regular Lagrangian,  we deduce:
\begin{itemize}
\item
 If $( \xi _1,\dots, \xi _k)$ is a
solution of (\ref{genericEL}) then it is a {\sc sopde}, (see
(\ref{locel3})).

 \item
  Since $ (\xi _1,\dots, \xi _k) $ is a {\sc sopde},
 from Proposition \ref{sope1} we know that, if it is integrable, then its integral
sections are first prolongations $\phi^{(1)}\colon\mathbb{R}^k\to
T^1_kQ$ of maps $\phi\colon \mathbb{R}^k \to Q$, and from
(\ref{locel4}) we deduce that $\phi$ is a solution to the
Euler-Lagrange equations (\ref{ELe}).
\item
 Equation (\ref{locel4}) leads us to define
local solutions to (\ref{genericEL}) in a neighborhood of each
point of $T^1_kQ$ and, using a partition of unity, global
solutions to (\ref{genericEL}).
 \item
In the case $k=1$,
 the equation (\ref{genericEL}) is $\imath_{\xi} \omega_L = dE_L$,
which is the dynamical equation of the Lagrangian formalism in
Mechanics.
\end{itemize}

\begin{example}
 Let $ L\colon   T^1_3\r \to \r$ be a Lagrangian given by
\begin{equation}\label{lag}L\colon   T^1_3\r \to \r, \quad L( q,v_1,v_2,v_3)= \displaystyle\frac{1}{2}(v_1^2-c^2(v_2^2+v_3^2)) \quad .
\end{equation}

Let us suppose that $(\xi_1,\xi_2,\xi_3)$ is a solution of the equation (\ref{genericEL})
$$ \sum_{A=1}^3 \inn_{\xi_A}\omega_L^A=\d E_L\, .$$
Since $L$ is regular we know that $(\xi_1,\xi_2,\xi_3)$  is  a {\sc sopde} satisfying  (\ref{locel4}). Then each $\xi_A$ is locally given by
$$
\xi_A= v_A  \displaystyle\frac{\partial }{\partial q }+(\xi_A)_1\displaystyle\frac{\partial }{\partial v_1 }+(\xi_A)_2\displaystyle\frac{\partial }{\partial v_2 }+(\xi_A)_3\displaystyle\frac{\partial }{\partial v_3 } \, , \quad 1\leq A \leq 3 \, .
$$
From  (\ref{locel4}) we have that
 \begin{equation}\label{kaka} 0=\frac{\partial^2 L}{\partial
v_A\partial v_B}( \xi _A)_B =   (\xi_1)_1-c^2 [(\xi_2)_2 +( \xi_3)_3] \end{equation}

  From    (\ref{nn1}) and (\ref{kaka})    we obtain that  if $\phi^{(1)}(\mathbf{t})$ is an integral section of the   $3$-vector field $(\xi_1,\xi_2,\xi_3)$ then $\phi:\rk \to Q$  satisfies the equation
$$
0=\displaystyle\frac{\partial^2 \phi}{\partial (t^1)^2 }- c^2(\displaystyle\frac{\partial^2 \phi }{\partial (t^2)^2 }+\displaystyle\frac{\partial^2 \phi }{\partial (t^3)^2 })$$
which is the {\it $2$-dimensional  wave equation}.\end{example}

\subsection{Linearizable {\sc sopde}s.}\protect\label{4.4}

In this section we introduce the definition of linearizable {\sc sopde} and we establish a necessary condition so that a {\sc sopde} is linearizable.

\begin{definition}\label{sopdelineal} A {\sc sopde} $\xi=(\xi_1,\ldots,\xi_k)$ on $\tkq$ is said to be
{\bf linearizable} if for each point on  $\tkq$, its components $(\xi_A)^j_B$ can be written as follows \begin{equation}\label{linsopde}
(\xi_A)^j_B=\big(\mathcal{A}^j_{AB}\big)^C_m\,v^m_C +
\big(\mathcal{B}^j_{AB}\big)_m\,q^m + \mathcal{C}^j_{AB} \,
\end{equation} with
$\big(\mathcal{A}^j_{AB}\big)^C_m,\,\big(\mathcal{B}^j_{AB}\big)_m,\,\mathcal{C}^j_{AB}\in\r$.
\end{definition}

\begin{prop}
If $\xi$ is linearizable then the curvature of the
nonlinear connection $\mathcal{H}_\xi$ vanishes.
\end{prop}

\begin{pf}
 Since $\xi$ is linearizable, from  (\ref{compgamxi}) and
(\ref{linsopde}) we obtain  that the  components of the nonlinear connection
   $\mathcal{H}_\xi$ are
$$
(N_\xi)^\imath_{Bj}=-\displaystyle\frac{1}{k+1}\displaystyle\sum_{A=1}^k
\big(\mathcal{A}^\imath_{AB}\big)^A_j\quad .
$$

Now from (\ref{loccur}) we deduce that the curvature $\Omega$ vanishes.
\end{pf}

   In the particular case of  a linearizable  {\sc sopde}, 
   Proposition \ref{sope1} can be written as follows.
\begin{prop} \label{sope*}
Let  $\xi=(\xi_1,\ldots,\xi_k)$ be a linearizable and integrable {\sc sopde}. If the first prolongation of $\phi:\rk\to Q$ is
    an integral section of $\xi=(\xi_1,\ldots,\xi_k)$
 then we have
\begin{equation}\label{nn10}
 \frac{\displaystyle\partial^2 \phi^j} {\displaystyle\partial t^A
\partial t^B       }(\mathbf{t})= (\xi_A)^j_B(\phi^{(1)}(\mathbf{t})) =
\big(\mathcal{A}^j_{AB}\big)^C_m\,\displaystyle\frac{\partial \phi^m}{\partial
t^C} + \big(\mathcal{B}^j_{AB}\big)_m\,\phi^m(\mathbf{t}) +
\mathcal{C}^j_{AB}
 \end{equation}
Conversely, if $\phi\,\colon \rk \to Q$ is a map satisfying
(\ref{nn10}), then $\phi^{(1)}$ is an integral section  of $\xi$.
\end{prop}

\begin{example} From (\ref{gamheat}) and (\ref{linsopde})  we deduce that
the {\sc sopde}   (\ref{gamheat}) is linearizable.\end{example}

\subsection*{Acknowledgments}

We acknowledge the partial financial support of {\sl Ministerio de
 Ciencia e Innovaci\'{o}n}, Project MTM2008-03606-E/  and  Project MTM2008-00689/MTM. We are grateful
to Prof. Ioan Bucataru ( ``Al.I.Cuza" University) by its suggestions and comments on this work.


\end{document}